\documentclass[usegraphicx,usenatbib,onecolumn]{mn2e}

\begin{document}

\title{The early reionization with the primordial magnetic fields}

\author[Tashiro, H. et al.]
{Hiroyuki Tashiro$^1$\
 and Naoshi Sugiyama$^2$\\
  $^1$Department of Physics, Kyoto University, Kyoto 606-8502, Japan\\
  $^2$Division of Theoretical Astronomy, National Astronomical Observatory,
 Japan, Mitaka, Tokyo 181-8588 Japan }

\date{\today}

%98.80.Es

\maketitle

\begin{abstract}

The early reionization of the intergalactic medium, which is favored
from the WMAP temperature-polarization cross-correlations, contests the
validity of the standard scenario of structure formation in the cold
dark matter cosmogony.  It is difficult to achieve early enough star
formation without rather extreme assumptions such as very high escape
fraction of ionizing photons from proto-galaxies or a top-heavy
initial mass function. Here we propose an alternative scenario that
is additional fluctuations on small scales induced by primordial
magnetic fields trigger the early structure formation.  We found that
ionizing photons from Population III stars formed in dark haloes can
easily reionize the universe by $z \simeq 15$ if the strength of
primordial magnetic fields is between $0.7$--$1.5 \times 10^{-9}$ Gauss.

\end{abstract}


\section{introduction}

The reionization process of the intergalactic medium (IGM) is one of
the major remaining problems in modern cosmology.  From the
Gunn-Peterson test of QSO absorption lines, it is known that the vast
majority of IGM is ionized by $z \sim 6$ \citep{becker, fan}. The
recent measurement of the cosmic microwave background (CMB)
temperature and polarization cross correlations by WMAP implies that
the optical depth of the universe is about $0.17$ \citep{spergel,
kogut}.  This result favors the early reionization scenario: the
reionization process occurs at $z=15$ -- $20$ and the reionization
sources are first stars, unlike quasars and galaxies which are known
as the reionization sources previously (for the details see
\citealt{early}).

The early reionization process by the stellar sources has been studied
in detail after WMAP~\citep{cen, fuku-kawa, c-f-w, s-l, h-h}.  In
these works, cold dark matter cosmogony with WMAP parameters is
employed.  What they found was it is difficult to get $\tau=0.17$ if
the standard Salpeter initial mass function (IMF) is adapted.  To have
early enough reionization, one needs to assume almost $100\%$ escape
fraction of ionizing photons from proto-galaxies, or introduce a
top-heavy IMF.  Heavy stars may form in the early universe induced by
the $\rm H_2$ molecular cooling while it is still little known about
the IMF in the early universe.

An alternative scenario to realize early reionization is to
enhance the amplitude of the CDM power spectrum on very small scales.
Such enhancement makes the dark haloes form earlier. Accordingly the
star formation process starts early enough to achieve $\tau=0.17$.
The observational data from redshift surveys of galaxies such as 2dF
and SDSS, and Ly-$\alpha$ clouds \citep{spergel, seljak} strongly
constrain the amplitude of the density fluctuations in the scales
larger than 1 Mpc.  However, the amplitude in the scales which are
relevant to the first star formation, $0.01$ -- $0.1$ Mpc, is still
unclear.  Therefore there is still room for considering additional
power in the power spectrum on very small scales.  For example, models
with initially running spectral index $n > 1$ on small scales,
additional fluctuations  from the isocurvature modes, or non-Gaussian
statistics are considered in the context of early
reionization~\citep{chen-cooray, avelino, s-z}.

If there exist strong enough primordial magnetic fields at the
reionization epoch, these magnetic fields produce additional density
fluctuations of baryons by the Lorentz force. %or the magnetic tension.
The magnetic tension is more effective on small scales where the
entanglements of magnetic fields are larger.  Therefore we expect to
have additional power in the density power spectrum on small scales as
is the case of isocurvature perturbations.  Structure formation by
magnetic fields was first discussed by \citet{wasserman}.
More detailed analysis was carried by \citet{k-o} and the
influence on the formation of large scale structure was recently
estimated by \citet{gopal}.  
\citet{sethi}
pointed out that nanoGauss magnetic fields can induce
early structure formation and may have the potential
to achieve the early reionization implied by the WMAP team. 
They also studied reionization of IGM induced by the dissipation of magnetic
energy. 
In this paper, we thoroughly
investigate the role of primordial magnetic fields on the early
reionization process. We concentrate on the early structure formation due to
the additional power spectrum generated by magnetic fields.

It is known that there exist magnetic fields of several $\mu$Gauss in
most of the galaxies and the clusters of galaxies while the origin of
these magnetic fields is still uncertain.  
Coherence lengths of these magnetic fields are typically $100$ kpc
--$1$ Mpc \citep{magobs}. 

Perhaps small seeds of the magnetic fields are produced inside
astronomical objects such as stars and AGNs due to the Biermann
battery mechanism.  Although the resultant magnetic fields are very
weak, those may be amplified by the dynamo process (\citealt{dynamo}; for
a comprehensive review see \citealt{magreview}).  Eventually these
magnetic fields are spread by Supernova winds or AGN jets into IGM.
However, 
for achieving observed values in
clusters of galaxies and high redshift galaxies~\citep{maghighz,
clustermag1, clustermag2},
there are difficulties in dynamo theory \citep{branden-subramanian}.

An alternative to the dynamo scenario is the generation of magnetic
fields in the early universe, which we consider in this paper.
Magnetic fields can be formed either due to the bubble collisions
during the cosmic phase transition such as QCD or electroweak phase
transitions, or due to the break of the conformal invariance in the
Maxwell theory during the inflation.  For a detailed review, see
\citealt{giova}.  These magnetic fields are formed
early enough to make influence on the first structure formation in the
universe and the reionization process.

The primordial magnetic fields are constrained by Big-Bang
nucleosynthesis and CMB
\citep{bbnconstraint1,bbnconstraint2,diss-j-k-o,mack,lewis,y-i,tashiro}.  The upper limit
of the comoving amplitude of magnetic fields is $\sim 10$ nGauss at the
$1$ Mpc scale.

% The primordial magnetic fields induce the density perturbations after
% the recombination \citep{wasserman,k-o,s-b-nonline,gopal,sethi}.  The
% density perturbations generated by the primordial magnetic fields have
% a blue spectrum.  From the large scale structure observations, it is
% allowed the magnetic fields with 1nGauss at 1Mpc exist at present
% \citep{gopal}.  The scales where first stars are formed is smaller than
% the scales of the large scale structures.  Therefore, there is still a
% possibility that the density fluctuations produced by the primordial
% magnetic fields dominate the primordial perturbations in such scales,
% without conflicting with the observations.  In this paper we evaluate
% the formation rate of the Population III stars from both the
% primordial density fluctuations and the density fluctuations induced
% by the magnetic fields through the Press-Schechter formalism and we
% assess the accomplishment of the early reionization in the simple
% reionization model.

Throughout this paper we use the cosmological parameters measured
by the WMAP teams,: the Hubble constant (in the unit of $100 ~{\rm km s}^{-1} {\rm Mpc}^{-1}$) 
$h= 0.71$, the matter density ratio $\Omega_{\rm m}= 0.27$ 
and the baryon density ratio $\Omega_{\rm b}= 0.044$ \citep{spergel}.

\section{the density fluctuations generated by the primordial magnetic fields}

Primordial magnetic fields produce the density perturbations after
recombination \citep{wasserman,k-o,s-b-nonline,gopal,sethi} by the
Lorentz force.  The generated perturbations grow gravitationally and
end up collapsing to form structure.  The density fluctuation
evolution with the primordial magnetic fields is governed by 
following equations,
\begin{equation}
{\partial^2 \delta_{\rm b} \over \partial t^2} = -2 {\dot a \over
a}{\partial \delta_{\rm b} \over \partial t } +4 \pi G (\rho _{\rm b}
\delta_{\rm b} + \rho _{\rm dm} \delta_{\rm dm} ) + S(t,{\bf x}),
\label{baryon-den-eq}
\end{equation}
\begin{equation}
S(t,{\bf x})={ \nabla \cdot \left[ (\nabla \times {\bf B}_{0} ({\bf
x})) \times {\bf B}_{0} ({\bf x})\right] \over 4 \pi \rho_{{\rm b} 0}
a^3 (t) },
\end{equation}
\begin{equation}
{\partial^2 \delta_{\rm dm} \over \partial t^2} = -2 {\dot a \over
a}{\partial \delta_{\rm dm} \over \partial t } +4 \pi G (\rho _{\rm b}
\delta_{\rm b} + \rho _{\rm dm} \delta_{\rm dm} ) ,
\label{dm-den-eq}
\end{equation}
where $a(t)$, $\delta$ and $\rho$ are the scale factor, 
the density perturbations and the energy density, and
the subscripts b, dm and $0$ denote the baryon and dark matter
components, and the present (comoving) value.
The dot represents the time derivative.

In Eq.~(\ref{baryon-den-eq}), we combine equations for ionized and
neutral baryons.  In the early universe, the interaction 
between ions and neutrals is so efficient that they are tightly
coupled and move together.  The interaction rate is proportional to 
the relative velocity between ions and neutrals.  Since we can assume
the thermal velocity of baryons as the relative velocity if there is
no magnetic fields, the interaction rate, which is also proportional
to the residual ionization fraction,  becomes shorter as the
universe expands.  Eventually, the time scale of the interaction
(inverse of the interaction rate) becomes longer than the Hubble time
at $z\approx 10$ for the value of the ionization fraction $x_e
\approx 10^{-4}$.  
Since then, we need to treat ions and neutrals
separately. However, if there exists magnetic fields, this is not the
case because the velocity of ions (and the relative velocity) 
becomes almost the Alfven velocity $v_{\rm A} \equiv B/\sqrt{4\pi
  \rho_{\rm b} x_e}$
which is $1000$ times larger than the thermal velocity at $z\approx
10$.  Accordingly the time scale of the interaction 
becomes $1000$ times shorter and 
never exceeds the
Hubble time before the reionization.

In order to solve these equations, we define total matter density and
its perturbation 
$\rho_{\rm m}$ and $\delta_{\rm m}$ as
\begin{equation}
\rho_{\rm m} =(\rho_{\rm b}+\rho_{\rm dm}),
\label{rhom-defi}
\end{equation}
\begin{equation}
\delta_{\rm m} ={ (\rho _{\rm b} \delta_{\rm b} + \rho _{\rm dm}
\delta_{\rm dm} ) \over \rho_{\rm m}}.
\label{deltam-defi}
\end{equation}
We can write the equations for $\delta_{\rm b}$ and $\delta_{\rm m}$ 
from Eqs.~(\ref{baryon-den-eq}) and (\ref{dm-den-eq}) as
\begin{equation}
{\partial^2  \delta_{\rm b} \over \partial t^2} 
= -2 {\dot a \over a}{\partial \delta_{\rm b} \over \partial t }
+4 \pi G \rho _{\rm m} \delta_{\rm m} 
+ S(t, {\bf x}),
\label{baryon-den-eq-2}
\end{equation}
\begin{equation}
{\partial^2 \delta_{\rm m} \over \partial t^2} = -2 {\dot a \over
a}{\partial \delta_{\rm m} \over \partial t } +4 \pi G \rho _{\rm m}
\delta_{\rm m} + {\rho_{\rm b} \over \rho_{\rm m}} S(t,{\bf x}).
\label{matter-den-eq-2}
\end{equation}
The solution of Eq.~(\ref{matter-den-eq-2}) can be acquired
by the Green's function method,
\begin{eqnarray}
\delta_{\rm m} =A({\bf x}) D_1 (t)+ B({\bf x}) D_2 (t) 
-D_1(t) {\Omega_{\rm b} \over \Omega_{\rm m}} \int^{t}_{t_{\rm i}} dt' {S(t' ,{\bf x}) D_2 (t') \over W(t')}
+D_2(t) {\Omega_{\rm b} \over \Omega_{\rm m}} \int^{t}_{t_{\rm i}} dt' {S(t' ,{\bf x}) D_1 (t') \over W(t')},
\label{solution-matter-eq}
\end{eqnarray}
where $D_1(t)$ and $D_2(t)$ are the homogeneous solutions of
Eq.~(\ref{matter-den-eq-2}), $W$ is the Wronskian and is expressed
as
\begin{equation}
W(t)=D_1(t) \dot D_2(t) -D_2(t) \dot D_1(t), 
\label{wronskian}
\end{equation}
and $t_{\rm i}$ denotes the initial time.  

The first and second terms of Eq.~(\ref{solution-matter-eq}) are
corresponding to the growing and the decaying mode solutions of
primordial perturbations produced by inflation and the third and
fourth terms are the ones generated by the primordial magnetic fields.
Hereafter we only consider the growing mode solution for the former
terms and describe it as $\delta_{\rm mP}$. The later two terms 
are written as $\delta_{\rm mM}$.  

Since we only treat the evolution of perturbations in the matter
dominated era, $D_1(t) \propto t^{2/3}$ and $D_2(t) \propto t^{-1}$
while they should be modified once the cosmological constant term
starts to dominate in expansion of the universe, $z \la 0.5$.  
Accordingly we obtain $\delta_{\rm mM} $ as~\citep{wasserman, k-o}
\begin{equation}  
\delta_{\rm mM} = {\Omega _{\rm b} \over \Omega _{\rm m}} \left[{9
\over 10} \left({t \over t _{\rm i} }\right)^{2/3} +{3 \over 5}
\left({t \over t _{\rm i} }\right)^{-1} -{3 \over 2}\right] t_{\rm i}
^2 S( t_{\rm i}, {\bf x}).
\label{mag-part}
\end{equation}
The density fluctuations generated by the primordial magnetic fields
have the same growth rate as the primordial density fluctuations.

Next, let us calculate the power spectrum of the matter density
perturbations.  For simplicity we assume that there is no correlation
between the magnetic fields and the primordial perturbations.  With
this assumption, the matter power spectrum can be described as
\begin{equation}
P_{\rm m}(k) = P_{\rm mP}(k)+P_{\rm mM}(k) 
\equiv \langle |\delta_{\rm mP}(k)|^2 \rangle + \langle |\delta_{\rm
  mM}(k)|^2 \rangle, 
\label{matter-power}
\end{equation}
where $\delta_{\rm mP}(k)$ and $\delta_{\rm mM}(k)$ are Fourier
components and 
$\langle ~ \rangle$ denotes the ensemble average.  

We obtain $P_{\rm mP}(k)$ by using the CMBfast code~\citep{cmbfast} while 
$P_{\rm mM}(k)$ is calculated from Eq.~(\ref{mag-part}) as 
\begin{equation}
P_{\rm mM}(k) = 
\left ( {\Omega _{\rm b} \over \Omega _{\rm m}} \right ) ^2
\left ( t_{\rm i}^2 \over 4 \pi \rho_{{\rm b}0}a^3 (t_{\rm i}) \right)^2
\left[{9 \over 10} \left({t \over t _{\rm i} }\right)^{2/3} 
+{3 \over 5} \left({t \over t _{\rm i} }\right)-{3 \over 2}\right]^2
I^2 (k),
\label{power-mag-part}
\end{equation}
where 
\begin{equation}
I^2 (k) \equiv \langle |\nabla \cdot (\nabla \times {\bf B}_0 ({\bf
x})) \times {\bf B}_0 ({\bf x})|^2 \rangle.
\label{nonline-convo}
\end{equation}

To calculate $I^2(k)$, we need to know the power spectrum of the 
magnetic fields.  The amplitude of the power spectrum $B_0^2(k)$ is
defined as 
\begin{equation}
\langle B_{0i}({\bf k_1}) B^* _{0j}({\bf k_2}) \rangle \equiv { (2 \pi)^3
\over 2 } \delta({\bf k_1- k_2}) \left(\delta_{ij}-{k_{1i} k_{2i}
\over k_1^2 } \right) B^2 _0 (k).  
\end{equation}
The nonlinear convolution Eq.~({\ref{nonline-convo}}) leads to 
\begin{equation}
I^2 (k) = \int dk_1 \int d \mu {B^2 _0 (k_1) B^2 _0 (|{\bf k - k_1}|)
\over |{\bf k - k_1}|^2 } [2 k^5 k_1^3 \mu+ k^4 k_1 ^4 (1-5 \mu^2) +
2k^3 k_1^5 \mu^3].  
\label{nonline-convo-2}
\end{equation}

For the power spectrum $B_0^2(k)$, we adopt the power law shape with 
sharp cutoff at $k_{\rm c}$ as 
\begin{equation}
4 \pi k^3 B^2 _0 (k) = (n+3) \left({k \over k_{\rm c}} \right) ^{n+3} B_0 ^2,
\qquad {\rm for}\ (k<k_{\rm c}), 
\label{power-B}
\end{equation}
where $n$ is the spectral index.
Here $B_0$ is the rms amplitude of the magnetic fields in real space
as $\langle B^2({\bf x}) \rangle = B_0 ^2$, which provides 
the characteristic magnetic amplitude at the cutoff scale.

The cutoff scale
depends on the dissipation mechanism of the magnetic field energy
\citep{j-k-o,s-b-nonline,b-j-paper,tashiro}.  Before recombination, 
dissipation is caused by the interaction between baryons and photons.
After recombination, the magnetic field energy in small scales is
mostly dissipated by the nonlinear effects, i.e., direct cascade.  The
dissipation time at the scale $l$ due to the direct cascade 
is the
eddy turn-over time, $l/v$, where $v$ is the baryon fluid velocity.
Since the interaction between baryons and photons which plays a role
as a viscosity is no longer effective, the velocity of the baryon fluids
increases and the equipartition between the magnetic fields and the
kinematic energy is established.  Hence, we assume that the fluid
velocity achieves the Alfven velocity.  
Under this assumption, the cutoff scale is
determined as the scale where the eddy turn-over time is equal to the
Hubble time $H^{-1}$, 
\begin{equation}
k_{\rm c} \approx 2 \pi { H \over v_{\rm A} } \approx 102 ~{\rm Mpc^{-1}} \left(
{B_0 \over 1~{\rm nGauss}} \right)^{-1} \left( {\Omega_{\rm b} \over 0.044}
\right)^{1/2} \left( {h \over 0.71} \right)^2.
\label{cutoff}
\end{equation}
Note that this comoving cutoff scale is constant in the matter
dominated epoch.  In this paper, we consider the power spectrum with
sharp cutoff below $1/k_{\rm c}$ for simplicity as is shown in
Eq.~(\ref{power-B}) to calculate $I^2(k)$.  With this assumption, $k_1$
integration of $I^2(k)$ is determined by the value of integrand at
$k=k_1$.  However, in reality, the direct cascade process of the
magnetic fields from large scales to small scales likely produces a
power-law tail below the cutoff scale in the power spectrum.  Although
most of the contribution for $I^2(k)$ still comes from the peak of the
spectrum, i.e,. $k_1=k_{\rm c}$ as well as the sharp cutoff case,
some corrections may be needed. 

Finally we introduce one more important scale for the evolution of
density perturbations, i.e., magnetic Jeans length.  
Below this scale, the magnetic pressure gradients, which we do not
take into account in Eq.~(\ref{matter-den-eq-2}), counteract the gravitational growth 
and prevent further growth of density perturbations. 
\citealt{k-o} evaluated the magnetic Jeans scale as 
\begin{equation}
k_{\rm MJ}=5 {\rho_{\rm m0} \sqrt{G} \over B_0}
= 32 ~{\rm Mpc}^{-1} \left({B_0 \over 1~{\rm nGauss}}\right)^{-1} 
\left({\Omega_{\rm m} \over 0.27}\right) \left({ h \over 0.71} \right)^2 .
\label{jeans}
\end{equation}

\section{reionization}

The Population III stars are considered as the main sources of 
reionization.  We can estimate the number of photons from Populations
III stars following \citealt{s-l}.  In their analysis,
ionizing photons are derived from the star-formation-rate density,
which is assumed to be proportional to the time derivative of the
fraction of the total mass in collapsed haloes $F$.  To obtain $F$,
the Press-Schechter prescription is employed, in which additional
small-scale power in the matter power spectrum induced by magnetic
fields plays an important role.

\subsection{The mass dispersion}

From the matter power spectrum $P_{\rm m}(k, t)$, the mass dispersion
within a radius $R$ can be written as 
\begin{equation}
\sigma^2 (R, t) = 4 \pi \int dk k^2 P_{\rm m} (k, t) \exp \left( -k^2 R^2 \right) .
\label{dispersion}
\end{equation}
Here we adopt the Gaussian window function. 

The total mass $M$ within
$R$ in the matter dominated epoch can be described as 
\begin{equation}
R=  18 ~ {\rm kpc}  \left({ M \over 10^6 M_\odot} \right) ^{1/3}
\left({ \Omega_{\rm m} \over 0.27}\right)^{-1/3} 
\left({ h \over 0.71}\right)^{-2/3}.
\label{radius-mass}
\end{equation}

\begin{figure}
  \begin{center}
    \includegraphics[keepaspectratio=true,height=50mm]{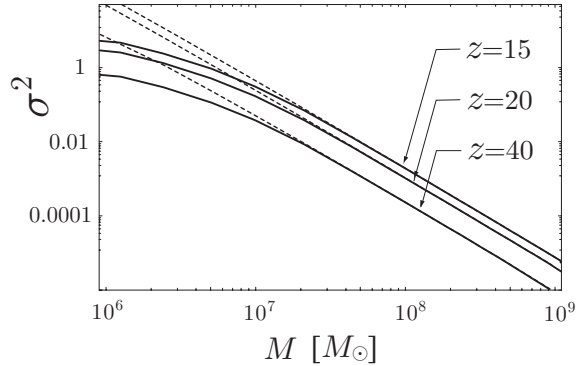}
  \end{center}
  \caption{The redshift evolution of the mass dispersion of the
  density perturbation generated by the magnetic fields with $B_0=0.7$
  nGauss and $n=1$.  The solid lines are the mass dispersion at
  $z=15$, $z=20$ and $z=40$ (from top to bottom).  
  The dotted lines are the mass dispersion without considering the
  effect of the magnetic Jeans oscillations.}
  \label{fig:sigma_mag}
\end{figure}

We plot the evolution of the mass dispersion which is calculated from the matter
power spectrum induced by magnetic fields in Fig.~\ref{fig:sigma_mag}.
On large mass scales, the mass dispersion is proportional to $
M^{-7/3}$ for the case of the spectral index $n=1$.  The dependence on
the spectral index is discussed below.  On the other hand, on the 
mass scales lower than the magnetic Jeans mass $M_{\rm MJ} \simeq 1.2 \times
10^{9} (B_0 /1~{\rm nGauss})^{3}M_\odot $, 
the slope of the mass dispersion
becomes milder due to the sharp cutoff of the matter power
spectrum below the magnetic Jeans scale in Fourier space.  
For the comparison, we also plot
the mass dispersion without taking into account the effect of the
magnetic Jeans oscillations as dotted lines.

\begin{figure}
  \begin{center}
    \includegraphics[keepaspectratio=true,height=50mm]{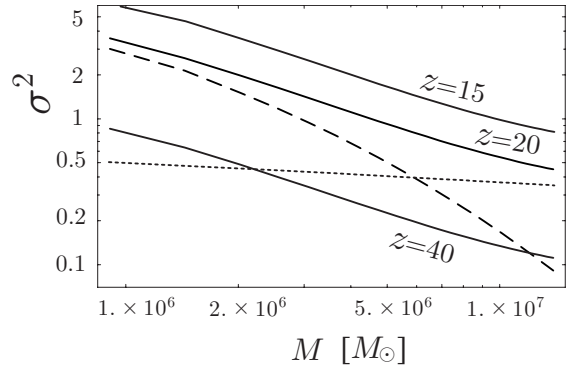}
  \end{center}
  \caption{The redshift evolution of the mass dispersion for the model
with $B_0=0.7$ nGauss and $n=1$.  The solid lines represent the total
mass dispersion at $z=15$, $z=20$ and $z=40$ (from top to bottom).
The dotted and dashed lines are the contributions from $P_{\rm mP}$
and $P_{\rm mM}$, respectively, at $z=20$.}
  \label{fig:sigma_redshift}
\end{figure}

Since the mass scales relevant to the Population III star formation
are around $10^6$ -- $10^7 ~M_\odot$ as is shown in the next
subsection, we hereafter focus on these mass scales.  In
Fig.~\ref{fig:sigma_redshift} the evolution of the mass dispersion is
seen for the model with $B_0=0.7$ nGauss and $n=1$.  Since the growth
rates of both $P_{\rm mP}$ and $P_{\rm mM}$ are $t^{4/3} \propto
1/(1+z)^2$, the mass dispersion $\sigma^2 \propto 1/(1+z)^2$ too.  In
this figure, we plot the contributions from $P_{\rm mP}$ and $P_{\rm
mM}$ separately at $z=20$.  It is shown that below $M=5.0 \times
10^{6}M_\odot$, the contribution from magnetic fields, i.e., $P_{\rm
mM}$ is dominated. This mass scale is independent on redshift.
Accordingly we expect to have early structure formation induced by
$P_{\rm mM}$.

\begin{figure}
  \begin{center}
    \includegraphics[keepaspectratio=true,height=50mm]{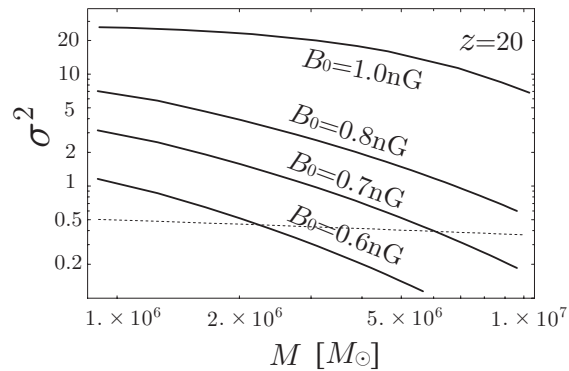}
  \end{center}
  \caption{ 
Mass dispersions of models with different strengths of the magnetic fields.  
The solid lines show the contributions from the magnetic fields
of $1.0$ nGauss, $0.8$ nGauss, $0.7$ nGauss and $0.6$ nGauss (from top to bottom) 
at $z=20$.  }
    \label{fig:sigma_strength}
\end{figure}

We show the mass dispersion for models with different magnetic field
amplitudes at $z=20$ in Fig.~\ref{fig:sigma_strength}.  The
contributions from magnetic fields are plotted as the solid lines.  The 
dotted line represents the contribution from the usual primordial perturbations.
We can analytically
estimate $B_0$ dependence from Eq.~(\ref{nonline-convo-2}) as
$\sigma^2 \propto I^2 \propto B_0^7$ which is consistent with
numerical results.  It is shown that the effect of magnetic fields on
early structure formation or reionization is significant if $B_0 \ga
0.6 $nGauss since there appears extra-power above the smallest
collapsed haloes to form Population III stars, i.e., $10^6$ -- $10^7 M_\odot$.
As the magnetic field strength becomes stronger, the magnetic Jeans
scale shifts to the large scale. 
Accordingly, the slope of the mass dispersion on the relevant mass
scales becomes flatter as is shown in  
Fig.~\ref{fig:sigma_strength}.

\begin{figure}
  \begin{center}
    \includegraphics[keepaspectratio=true,height=50mm]{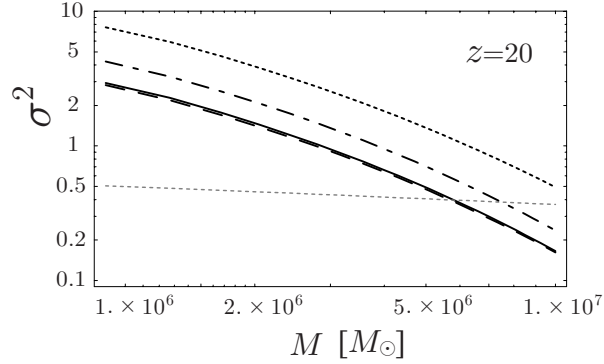}
  \end{center}
  \caption{
Mass dispersions of models with different spectral indices for the
power spectra of the magnetic fields at $z=20$. 
We fix the strength of the magnetic fields as $0.7$ nGauss.  
The dotted, dotted-dashed, solid, dashed lines 
(from top to bottom) are the models with 
$n=-2$, $-1$, $1$ and $2$, respectively.  
The thin dotted line is the contribution from the primordial perturbations.
}
    \label{fig:sigma_spectral}
\end{figure}

The mass dispersions for various spectral indices at $z=20$ are
plotted in Fig.~\ref{fig:sigma_spectral}.  The dependence on $n$ can
be explained as follows.  In the limit of $k/k_c \ll1$, 
$I^2(k) \sim \alpha k^{2n+7} + \beta  k^{4}$
%$I^2(k) \sim \alpha k^{2n+7} + \beta k_{\rm c} ^{2n+3} k^{4}$ 
where $\alpha$ and
$\beta$ are coefficients which depend on $n$ and $B_0$ \citep{k-o,
gopal}.  The former term dominates if $n<-1.5$, while the later one
dominates for $n>-1.5$.  From the definition Eq.~(\ref{dispersion}),
the mass dispersion $\sigma^2 \propto k^3 P_{\rm m}(k) \propto k^3 I^2(k)$.
Accordingly $\sigma^2 \propto k^{(3+4)} \propto M^{-7/3}$ for $n>-1.5$
and $k^{3+ (2n+7)} \propto M^{-(2 n+10)/3}$ for $n<-1.5$.  It is
interesting that the slope of $\sigma^2$ does not depend on $n$ for
$n>-1.5$.  Note that the slope becomes milder in the small mass
around the Jeans scale regardless of the value of $n$ as mentioned above.

\subsection{The ionization photon number}

We assume that the Population III stars are formed in collapsed haloes
with masses larger than $M_{\rm crit}$ and lower than $M_{\rm vir}$.
Here we adopt $M_{\rm crit}=1.0 \times 10^6 ~h^{-1} M_\odot $ and
$M_{\rm vir}=M(T_{\rm vir} = 10^4 {\rm K})$ where $M(T_{\rm vir})$ is
the virial mass with temperature $T_{\rm vir}$.  The virial mass
$M(T_{\rm vir})$ evolves as $(1+z)^{-1}$ for given temperature $T_{\rm vir}$.  
At $z=15$, the  virial mass with $T_{\rm vir}=10^4$ is $M(T_{\rm
vir}=10^4)=5.2 \times 10^6 M_{\odot}$.
 
Following the Press-Schechter prescription, we can derive the fraction
of the collapse haloes with masses larger than $M$ at time $t$ as
\begin{equation}
F(>M,t) = 1- \int^{ \delta_{\rm c}} _0 dx { \sqrt{2} \over \sqrt{\pi} \sigma(M,t) } 
\exp \left(-{ 1 \over 2 }{ x^2 \over \sigma^2 (M,t) } \right), 
\label{p-s-formalism} 
\end{equation}
where $\sigma^2 (M,t)$ is obtained by substituting
Eq.~(\ref{radius-mass}) into Eq.~(\ref{dispersion}).

From $F$, the global star-formation-rate density of the Population III
stars can be calculated as
\begin{equation}
\dot \rho_* (t) = e_* \rho_{\rm b} { d \over dt} \left[F(>M_{\rm
crit},t)-F(>M_{\rm vir},t) \right],
\end{equation}
where $e_*$ is the efficiency of conversion of gas into stars and
we adopt $e_*=0.001$~\citep{yoshida-abel}. 

Now we assume that the lifetime of Population III stars is
$\Delta t= 3 \times 10^6$ yr. and the production rate of ionizing
photons by Population III stars is $N_{\gamma,0}= 1.6 \times 10^{48}$
photons $s^{-1} M ^{-1} _\odot$ \citep{bromm}.
%In this case 
Accordingly 
we obtain the total production rate of ionizing photons as
\begin{eqnarray}
\dot n_{\gamma} (t) &=& \int^{t} _{t_{\rm i}}d t' N_{\gamma,0}
\theta(\Delta t -(t-t')) \dot \rho_* (t') \nonumber \\ &=&
N_{\gamma,0} \left[ \rho_* (t) - \rho_* (t-\Delta t) \right],
\label{cumu-rate}
\end{eqnarray}
where $\theta (t)$ is a step function
which is $\theta (t)=1$ for $t \ge 0$ and $\theta (t)=0$ for $t<0$. 

The cumulative number of photons per H atom is a good indicator of the
ionization degree of IGM.  It is known that about 20 cumulative
photons per H atom are requited in order to achieve a volume weighted
ionization of 99 percent \citep{h-a-m, s-a,s-y}.  From
Eq.~(\ref{cumu-rate}), the cumulative photons per H atom at time $t$
is represented as
\begin{eqnarray}
{n^{\rm cumul} _\gamma \over n_{\rm H} }(t) &=& {\mu m_{\rm p} \over
\rho_{\rm b}} \int^{t} _0 dt' \dot n_\gamma (t') \nonumber\\ &=&
N_{\gamma,0} {\mu m_{\rm p} \over \rho_{\rm b}} \int^{t} _0 dt' \left[
\rho_* (t') - \rho_* (t'-\Delta t) \right],
\end{eqnarray}
where $n_{\rm H}$ is the Hydrogen number density, $\mu$ is the mean
molecular weight, and $m_{\rm p}$
is the proton mass. 

\begin{figure}
  \begin{center}
    \includegraphics[keepaspectratio=true,height=50mm]{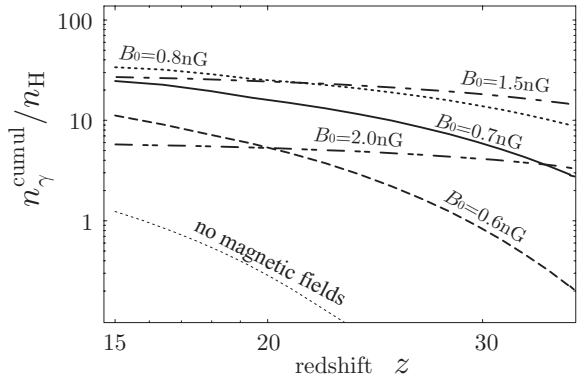}
  \end{center}
  \caption{ The cumulative photons from the Population III stars per H
  atom for the models with the magnetic fields.  The spectral index of
  the magnetic fields is fixed as $n=1$.  The strengths of the
  magnetic fields are $2.0$, $1.5$, $0.8$, $0.7$ and $0.6 $ nGauss for
  the dashed--dotted-dotted, dashed--dotted, dotted, solid and dashed
  lines, respectively.  The thin dotted line is
  the case with no magnetic fields.  }
  \label{fig:reion_strength}
\end{figure}

We show cumulative photons from the Population III stars per H
atom as a function of redshift.  
In Fig.~\ref{fig:reion_strength},
the cumulative photons are plotted for
different values of the magnetic strength $B_0$.  Here we fix $n=1$.
It is found that more photons are produced 
as the magnetic field strength becomes stronger since the amplitude of
the density perturbations depends on the strenth. 
We can conclude that the universe is reionized early enough to be
consistent with WMAP data if 
$B_0$ is larger than $0.7$ nGauss. 
Remember that 20 cumulative photons
are needed to achieve a 99 percent volume weighted ionization.

If the magnetic field strength is larger than $1$nGauss, however, it is
shown that the cumulative photons become smaller as we increase the
strength since the magnetic Jeans scale shifts to the larger scale
which makes Population III stars difficult to be formed.   
We found that the magnetic fields larger than $1.5$ nGauss
cannot generate large enough number of cumulative photons to  
induce early reionization.  

\begin{figure}
  \begin{center}
    \includegraphics[keepaspectratio=true,height=50mm]{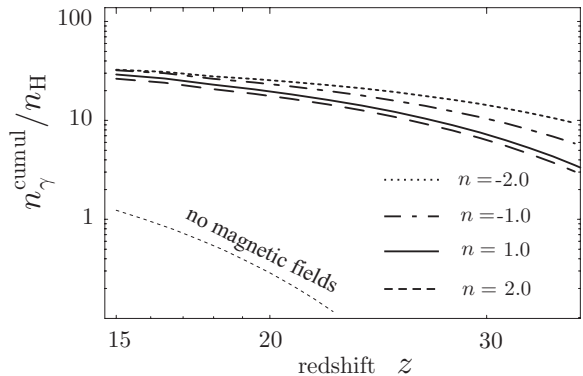}
  \end{center}
  \caption{ The cumulative photons from the Population III stars per H
atom for the models with the fixed strength of magnetic fields
$0.7$ nGauss and various spectral indices.  
The dotted, dotted-dashed, solid, dashed lines 
(from top to bottom) are the models with 
$n=-2$, $-1$, $1$ and $2$, respectively.  
It is shown that dependence on $n$ of the 
cumulative photons is very weak.  
The thin dotted line is for the case of no magnetic fields.  
}
  \label{fig:reion_spectral}
\end{figure}

Fig.~\ref{fig:reion_spectral} shows the dependence of 
the cumulative photons on $n$.  It is found that there is little
dependence if $n>-1.5$.  Therefore 
the requirement $B_0 \ga 0.7$ nGauss for early reionization is
robust regardless of the power law index of the magnetic fields.

\section{conclusion}

In this paper we investigate the role of the additional density
perturbations generated by the primordial magnetic fields on the
reionization process in the early universe.  These additional density
perturbations may trigger the early structure and star formation.
Employing a simple analytic recipe, we estimate the number of ionizing
photons emitted from the Population III stars.  We found that the
reionization process almost completes by $z \sim 15$ if the strength
of primordial magnetic fields is larger than $0.7$ nGauss and less
than $1.5$ nGauss.  Note that we adopt the Gaussian window function to
calculate the mass dispersion~(see Eq.~(\ref{dispersion})).  Different
choice of the window function alters the mass dispersion at the
magnetic Jeans scale.   Accordingly the magnetic field strength to be
required for the early enough reionization is also changed.  
If we employ the sharp-$k$ window function, for example, the
strength should be between $0.5$ and $0.7$ nGauss.

Such magnetic fields are not yet ruled out from current observations,
i.e., BBN, CMB temperature anisotropies and polarization, and Faraday
rotation of polarized lights from radio sources.  Although the
formation process of such primordial magnetic fields is still
uncertain, magnetic fields may be naturally generated during the
cosmological phase transition \citep{b-j-paper}.

The extra-power of the matter power spectrum will be directly
probed by future observations such as the fluctuations of the Hydrogen
21cm line \citep{loeb-zaldarriaga} and the substructure of lensing haloes
\citep{dalal-kochanek}.  Moreover, the thermal diffusion process of primordial
magnetic fields may cause ionization of IGM even before Population III
stars.  A measurement of fluctuations of the 21cm line will be
a powerful tool to investigate such pre-reionization~\citep{h-s-21}.

\section*{Acknowledgements}
We would like to thank an anonymous referee for some useful comments. 
N.S. thanks Carlos Cunha for his mention about evolution of density
perturbations induced by primordial magnetic fields.  
N.S. is supported by a Grant-in-Aid for Scientific Research from the
Japanese Ministry of Education (No. 17540276).

%It is true that 
%the early reionization scenario favored by the WMAP data 
%does not support the presence of the primordial magnetic fields.
%In this paper we discuss the possibility of the early reionization by the primordial magnetic fields
%in the very simple model.
%However, there are many uncertain about the effects of the primordial magnetic fields on the first star formation.
%[For example, the presence of the primordial magnetic fields causes the MRI dynamo instability in the accretion disk 
%and the magnetic fields rapidly amplified.
%Consequently, the magnetically driven ejection may make the accretion efficiency lower \citep{silk}. ]

%@ARTICLE{cmbfast,
%   author = {{http://www.cmbfast.org/}, },
%}

%\bibliography{magreion}

\end{document}